\documentclass[12pt,psf,epsf]{article}
\usepackage[utf8]{inputenc}
\usepackage[english]{babel}
\pdfoutput=1

\usepackage[section]{placeins}

\usepackage{standalone}
\usepackage{ulem}
\usepackage{changepage}
\usepackage{authblk}
\usepackage[centertags]{amsmath}
\usepackage{amssymb}
\usepackage[]{graphicx}
\usepackage{epsfig}
\usepackage{array}
\usepackage{amsthm}
\usepackage{latexsym}
\usepackage[mathcal]{euscript}
\usepackage{jheppub}
\usepackage{hyperref}

\usepackage{psfrag}
\usepackage{float}
\usepackage{cancel}
\usepackage{mathrsfs}
\usepackage{amsfonts}
\usepackage{amsmath}
\usepackage{slashed}
\usepackage{bm}
\usepackage{color}
\usepackage{subcaption}
\usepackage{amsmath}

\newcommand{\be}{\begin{equation}}
\newcommand{\ee}{\end{equation}}
\newcommand{\bea}{\begin{eqnarray}}
\newcommand{\eea}{\end{eqnarray}}

\newcommand{\RNumb}[1]{\textbf{\uppercase\expandafter{\romannumeral #1\relax}}}
\newcommand{\nn}{\nonumber}

\newcommand{\Tr}{\mathop{\mathrm{Tr}}}

\newcommand{\m}{\text{m}}
\newcommand{\gen}{\text{gen}}

\newcommand{\extgen}{^\text{ext}_\text{gen}}

\usepackage{soul}
\usepackage{comment}
\usepackage{xcolor}

\definecolor{darkblue}{rgb}{0,0,1}

\definecolor{dgreen}{rgb}{0,0.6,0}

\definecolor{darkraspberry}{rgb}{0.9,0.,0.3}

\definecolor{ddgreen}{rgb}{0,0.8,0}

\usepackage{comment}

\makeatletter
\gdef\@fpheader{\mbox{}} 
\makeatother

\begin{document}


\title{Entanglement Entropy in Jackiw–Teitelboim de Sitter gravity with Timelike Boundaries}

\author{Timofei Rusalev}

\affiliation{Steklov Mathematical Institute, Russian Academy of Sciences, Gubkin str. 8, 119991 Moscow, Russian Federation}

\emailAdd{rusalev@mi-ras.ru}

\abstract{The consideration of timelike boundaries in de Sitter static patches has a broad motivation, such as the formulation of a well-defined canonical ensemble and the realization of a natural framework for static-patch holography. In this work we study Jackiw–Teitelboim de Sitter gravity with symmetric timelike reflecting boundaries, which, in the presence of both cosmological and ``black hole'' horizons, naturally separate the spacetime into a ``black hole system'' and a ``cosmological system''. We apply the island formula to compute the entanglement entropy of conformal matter in both systems. In the “black hole system’’ an island appears, causing the entanglement entropy to saturate at the horizon value and preventing late-time growth. In the ``cosmological system’’ no island appears, and the entanglement entropy can become arbitrarily large depending on the position of the boundaries, indicating a tension with unitarity.}

\maketitle

\clearpage

\section{Introduction}

On cosmological scales the Universe is well approximated by de Sitter spacetime, a maximally symmetric geometry with positive cosmological constant that also describes the inflationary epoch in its earliest moments. Due to accelerated expansion, a static observer can access only a limited causal region bounded by a cosmological horizon~\cite{Spradlin:2001pw}. This horizon closely parallels black hole event horizons: both have a characteristic temperature and an entropy proportional to the horizon area, reflecting thermal radiation~\cite{Gibbons:1977mu}. In the de Sitter case the radiation is the Gibbons–Hawking effect, and the horizon carries the corresponding entropy. Despite these analogies, the thermodynamics of de Sitter horizons and their microscopic interpretation remain more challenging than in the black hole case, because the cosmological horizon is intrinsically observer-dependent \cite{Witten:2001kn, Bousso:2002fq, Goheer:2002vf, Anninos:2012qw}.

One of the most important challenges concerning black holes and their horizons is the information paradox \cite{Hawking:1975vcx, Hawking:1976ra}. Semiclassical analyses of Hawking radiation seem to imply that the entanglement entropy of the radiation, due to its thermal nature, increases monotonically and ultimately surpasses the horizon entropy \cite{Hawking:1976ra}. Such behavior implies a transition from a pure to a mixed state and thus a violation of unitarity, which is referred to as the black hole information paradox. The unitary scenario, however, is described by the Page curve~\cite{Page:1993wv, Page:2013dx}: for evaporating black holes the entropy first increases and then returns to zero, while for eternal black holes it rises and saturates at a value of order the thermodynamic entropy of the horizon.
 
Significant progress in resolving the information paradox has been achieved through the quantum extremal surface prescription \cite{Engelhardt:2014gca, Penington:2019npb, Almheiri:2019psf, Almheiri:2019hni}, developed in the context of holography, and the island formula, derived via the gravitational replica method in two-dimensional dilaton gravity models \cite{Penington:2019kki, Almheiri:2019qdq}, most notably in Jackiw–Teitelboim (JT) gravity in AdS. According to these results, the fine-grained entropy of a gravitational system is obtained by extremizing the generalized entropy, defined as the sum of the area term and the matter contribution. It is generally believed that the entanglement entropy computed in this framework reproduces the Page curve and is therefore compatible with unitary evolution.

Since the cosmological horizon is likewise associated with Gibbons–Hawking entropy and thermal radiation, one can formally study the time evolution of the entanglement entropy of this radiation and its consistency with unitarity. The study of islands in cosmological backgrounds has been the subject of numerous works \cite{Chen:2020tes, Hartman:2020khs, Sybesma:2020fxg, Balasubramanian:2020xqf, Geng:2021wcq, Aalsma:2021bit, Kames-King:2021etp, Teresi:2021qff, Seo:2022ezk, Levine:2022wos, Ageev:2023mzu}.

There exist spacetimes that contain both a black hole and a cosmological horizons. The Schwarzschild–de Sitter (SdS) spacetime is the canonical example. In two dimensions, this structure arises in the JT de Sitter gravity \cite{Maldacena:2019cbz, Cotler:2019nbi, Arefeva:2019buu, Svesko:2022txo, Moitra:2022glw, Nanda:2023wne, Cotler:2024xzz, Batra:2024qju, Held:2024rmg}. There exist two versions of JT de Sitter gravity, the full reduction and half reduction models \cite{Svesko:2022txo}. In this paper, we focus on the former, as it features both horizons and corresponds to the reduction of SdS in the Nariai limit. In SdS the coexistence of horizons implies particle creation at two distinct temperatures, associated respectively with the black hole and cosmological horizons, which complicates the analysis of their thermodynamic properties \cite{Shankaranarayanan:2003ya, Choudhury:2004ph, Pappas:2017kam, He:2018zrx, Singha:2021dxe}. For this reason it is common to introduce timelike boundaries between the horizons, which partition the spacetime into isolated “black hole” and “cosmological” systems~\cite{Gomberoff:2003ea, Sekiwa:2006qj, Saida:2009ss, Ma:2016arz, Bhattacharya:2021zgd}. Besides SdS, the introduction of timelike boundaries is also widely used in various spacetimes with a positive cosmological constant, including pure de Sitter space and JT de Sitter gravity, to study entanglement entropy \cite{Goswami:2022ylc, Yadav:2022jib, RoyChowdhury:2023eol}, define canonical ensembles~\cite{Svesko:2022txo, Banihashemi:2022jys}, formulate static-patch holographic dualities \cite{Parikh:2004wh, Banks:2005bm, Anninos:2011af, Anninos:2017hhn, Leuven:2018ejp, Susskind:2021omt, Susskind:2021dfc, Susskind:2021esx, Shaghoulian:2021cef, Shaghoulian:2022fop}, and investigate energy conditions~\cite{Batra:2024qju}.

In this work we study, within the framework of the island formula, the time evolution of entanglement entropy in both the black hole and cosmological systems of the full reduction model of JT de Sitter gravity, obtained by introducing boundaries. Specifically, we place timelike boundaries symmetrically in the two static patches and impose perfectly reflecting boundary conditions for the conformal matter fields. The entangling regions are composed of two intervals, one in each static patch, extending from a point near the black hole or cosmological horizon to the corresponding boundary. For the entanglement entropy calculations we employ BCFT$_2$ techniques using correlators of twist operators in curved two-dimensional spacetimes with a boundary, following the approach of \cite{Ageev:2023hxe} developed for Schwarzschild black holes with reflecting boundaries. For a simply connected region, this approach reproduces the entanglement entropy obtained in \cite{Franken:2023ugu} for the half-reduction model of JT de Sitter gravity, which naturally includes boundaries and in which the calculation was performed without twist operators.

Adopting the terminology used in studies of the Page curve for eternal black holes~\cite{Almheiri:2019yqk}, we refer to an ``information paradox” as the situation where the entanglement entropy of the radiation exceeds the thermodynamic entropy of the corresponding horizon in a given system. We show that in both the black hole and cosmological systems an information paradox arises in the absence of islands, when the boundary is located near the horizon opposite to that defining the system (e.g. near the cosmological horizon in the black hole system). Our main results can be summarized as follows:
\begin{itemize}
\item \textbf{Black hole system}. We find an island configuration that resolves the information paradox, in close analogy with the simply connected islands considered for eternal black holes \cite{Almheiri:2019yqk}. The difference is that in the present setup the island is located near the endpoints of the entanglement region rather than near the horizon.

\item \textbf{Cosmological system}. We show that an island solution does not arise, which may indicate a breakdown of unitarity. A similar conclusion was reached previously for the half reduction model of JT de Sitter gravity, where the absence of island was demonstrated \cite{Ageev:2023mzu}.
\end{itemize}

It is natural at this point to compare our results with the so-called “blinking island” effect observed for the two-sided Schwarzschild black hole with a spherically symmetric reflecting boundary \cite{Ageev:2023hxe}. This effect consists in the disappearance of the island for a finite interval of time, which, when the boundary is placed sufficiently far from the horizon, leads to a breakdown of unitary evolution. In the black hole system in JT de Sitter gravity, which parallels the Schwarzschild setup with reflecting boundaries, the island exists for all times (i.e. the blinking island effect is absent) and thereby resolves the information paradox. In the Schwarzschild setup with reflecting boundaries the heat capacity changes sign depending on the boundary position \cite{York:1986it}, whereas in the JT black hole system it remains positive for any boundary location \cite{Svesko:2022txo}. In this work we do not examine the relation between thermodynamic and entanglement properties, although this may be of interest for future studies.

Scenarios where the island configuration fails to yield unitary evolution, as in the case of the cosmological system in JT gravity, have been analyzed in \cite{Ageev:2023mzu, Ageev:2023hxe, Arefeva:2021kfx, Ageev:2022hqc, Ageev:2022qxv}. Of particular interest is the development of the idea of the principle of maximum entropy, investigated in \cite{Volovich:2023vib}, for the case of JT de Sitter gravity.
\\

The paper is organized as follows. In Section \ref{sec:setup}, we setup the notation, present JT de Sitter gravity with boundaries and the algorithm for entanglement entropy calculation. Section \ref{sec:bh} is devoted to the   entanglement entropy dynamics in black hole system. Then in Section \ref{sec:cosm}  similarly we study entanglement in cosmological system. Final section is devoted to conclusions.

\section{Setup}\label{sec:setup}

\subsection{JT de Sitter gravity}

The action for JT dilaton gravity with a positive cosmological constant $\Lambda > 0$ is%
\be\label{eq:IJT}
  I_{\text{JT}}=\frac{1}{16\pi G_2}\int_{M}d^2x\sqrt{-g}\left((\phi_0+\phi)R-2\Lambda\phi\right)
        +\frac{1}{8\pi G_2}\int_{\partial M}dy\sqrt{-h}\,(\phi_0+\phi)K.
\ee
This action can be obtained by spherical dimensional reduction of the \(d\)-dimensional Schwarzschild–de Sitter geometry, where $d \geq 4$, in the Nariai limit \cite{Svesko:2022txo}. The dilaton decomposes as \(\Phi=\phi_0+\phi(x)\), where the constant part \(\phi_0\) is proportional to the entropy of the higher-dimensional Nariai black hole, while the dynamical field \(\phi(x)\) describes deviations away from this limit. Here \(G_2\) denotes the dimensionless two-dimensional Newton constant.

The action \eqref{eq:IJT} describes the full reduction version of JT de Sitter gravity. There also exists the half reduction model, obtained by dimensional reduction of three-dimensional de Sitter space \cite{Svesko:2022txo}, whose action corresponds to \(\phi_0=0\).

The classical equations of motion for the metric and the dilaton in the absence of matter fields, which follow from \eqref{eq:IJT}, are given by
\bea
\label{eq:classic-equation-for-metric}
g_{\mu\nu}\nabla^2 \phi -\nabla_{\mu}\nabla_{\nu} \phi+g_{\mu\nu}\Lambda \phi &=&0, \\
R - 2\Lambda&=& 0. \label{eq:classic-equation-for-dilaton}
\eea

\begin{figure}[h!]\centering
    \includegraphics[width=0.8\textwidth]{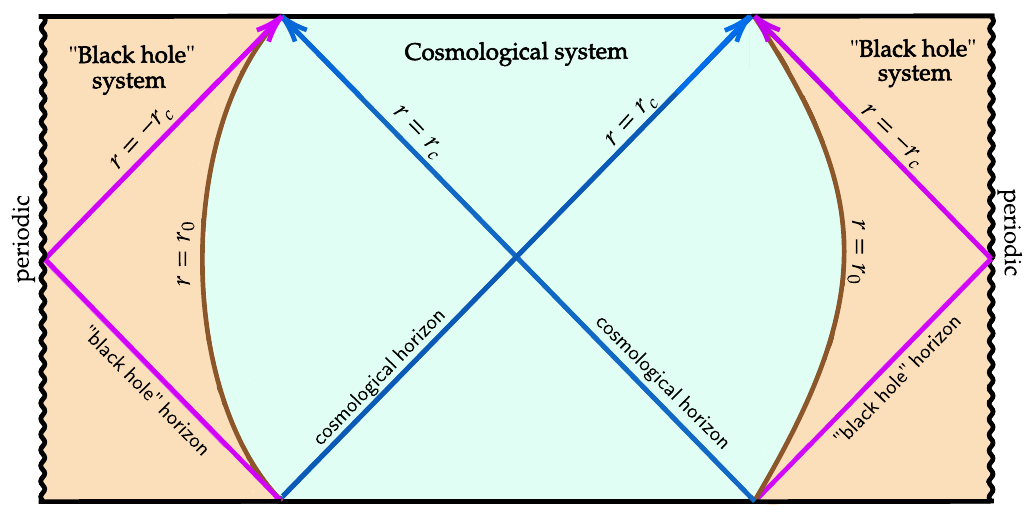}
	\caption{Penrose diagram for full reduction model geometry, which is similar to the Penrose diagram for the higher-dimensional Schwarzschild-de Sitter In each static patch there is a boundary at $r = r_0$, which, due to the periodic identifications (indicated by the wavy lines), divides the system into the cosmological system (shown in blue) and the black hole system (shown in brown).}
    \label{fig:caus-struct}
\end{figure}

Equation \eqref{eq:classic-equation-for-dilaton} implies that the geometry is locally two-dimensional de Sitter spacetime. 
In static coordinates \((t,r)\) the solution takes the form
\be\label{eq:static-sol}
ds^2 = - f(r)\,dt^{2} + \frac{dr^2}{f(r)}, 
\qquad f(r) = 1 - \frac{r^2}{r_c^2}, 
\qquad \Phi(r) = \phi_0+\phi_r \frac{r}{r_c},
\ee
with $r_c = 1/\sqrt{\Lambda}$. In two-dimensional de Sitter space the radial coordinate extends to negative values, so $r \in (-r_c, r_c)$.  The function \(f(r)\) vanishes at \(r=\pm r_c\), so that the geometry contains two horizons: the cosmological one at \(r=r_c\) and the ``black hole'' one at \(r=-r_c\). 
The constant $\phi_r$ is taken to satisfy $\phi_r \ll \phi_0$ so as to remain close to the Nariai limit, within which the action \eqref{eq:IJT} has been derived \cite{Svesko:2022txo}. The causal structure of considered two-dimensional space-time, depicted in Fig.\ref{fig:caus-struct}, is related to the higher-dimensional Schwarzschild-de Sitter spacetime. In two-dimensional de Sitter spacetime, the Cauchy surface has the topology of a circle, which implies a periodic identification on the Penrose diagram.

Let us comment on the presence of a ``black hole'' horizon. The full reduction model inherits structure from the higher-dimensional Schwarzschild–de Sitter spacetime, which contains both a black hole horizon and a cosmological horizon. Accordingly, the surface \(r=-r_c\) is identified with the black hole horizon. Also, in dilaton gravity the vanishing of the dilaton, \(\Phi = 0\), which is inversely proportional to the ``effective Newton coupling'', is interpreted as a singularity [link], as a particle can reach the surface \(\Phi=0\) in finite affine parameter. However, the local geometry at every point remains that of two-dimensional de Sitter space with constant curvature. So, we refer to $r=-r_c$ as the black hole horizon, following the convention in the literature \cite{Svesko:2022txo}.

It is convenient to introduce the ``cosmological'' Kruskal coordinates
\be\label{eq:cosm-krusk}
\tilde{U} = -\frac{1}{\kappa_c} e^{\kappa_c (t-r_*(r))},\qquad
\tilde{V} =  \frac{1}{\kappa_c} e^{-\kappa_c (t+r_*(r))},
\ee
where \(\kappa_c = 1/r_c\) is the surface gravity, and the tortoise coordinate $r_*(r)$ is defined by
\be
    r_{*}(r) = \frac{r_c}{2}\,\log\!\frac{r_c + r}{|r_c - r|}
\ee
with asymptotic behavior \(r_*(r)\to +\infty\) at \(r\to r_c\) 
and \(r_*(r)\to -\infty\) at \(r\to -r_c\). Solution \eqref{eq:static-sol} can be rewritten as
\be\label{eq:metric-cosmol}
ds^2 = -\frac{4\,d\tilde{U}\,d\tilde{V}}{\bigl(1-\kappa_c^{2}\tilde{U}\tilde{V}\bigr)^2},\qquad
\tilde{\Phi}(\tilde{U},\tilde{V})=\phi_0+\phi_r\frac{1+\kappa_c^{2}\tilde{U} \tilde{V}}{1-\kappa_c^{2}\tilde{U}\tilde{V}}.
\ee
These coordinates are regular at the cosmological horizon \(r=r_c\) and become singular at \(r=-r_c\). 
They can be analytically extended through the cosmological horizon into the region 
\(\tilde{U}>0,\,\tilde{V}<0\), corresponding to the left static patch, 
as well as into the regions with \(\tilde{U}>0,\,\tilde{V}>0\) and \(\tilde{U}<0,\,\tilde{V}<0\), 
which describe the domain \(r>r_c\).

In complete analogy, we introduce the ``black hole'' Kruskal coordinates
\be\label{eq:bh-krusk}
U =  \frac{1}{\kappa_c} e^{-\kappa_c (t-r_*(r))},\qquad
V = -\frac{1}{\kappa_c} e^{\ \kappa_c (t+r_*(r))},
\ee
for which
\be\label{eq:metric-bh}
ds^2 = -\frac{4\,dU\,dV}{\bigl(1-\kappa_c^{2}UV\bigr)^2},\qquad
\hat{\Phi}(U,V)=\phi_0-\phi_r\frac{1+\kappa_c^{2}U V}{1-\kappa_c^{2}U V}.
\ee
This chart is regular at \(r=-r_c\) and singular at \(r=r_c\), and admits analytic extension beyond the black hole horizon. We emphasize the relative sign difference in the dilaton expressions in \eqref{eq:metric-cosmol} and \eqref{eq:metric-bh}, so the identity $\Phi(r)=\tilde{\Phi}(\tilde{U}\tilde{V}(r))=\hat{\Phi}(U V(r))$ holds.

\subsection{Geometry with boundary and its Euclidean version}

We introduce two timelike spherically symmetric boundaries at the radial coordinate $r = r_0$ between black hole and cosmological horizons, i.e. at $r_0 \in (-r_c, r_c)$, in both static patches, see Fig.\ref{fig:caus-struct}. Introducing two boundaries leads to a natural decomposition of the full reduction geometry into two disconnected systems:
\begin{itemize}
    \item \textbf{Cosmological system:} the domain containing the cosmological horizon $r = r_c$, both static patches restricted to $r_0<r<r_c$, as well as the future and past domains beyond the cosmological horizon at $r>r_c$. This system is entirely covered by the cosmological Kruskal coordinates \eqref{eq:cosm-krusk}, which motivates their name.
    \item \textbf{Black hole system:} the complementary domain, which contains the black hole horizon $r = -r_c$, the remaining parts of both static patches with $-r_c < r < r_0$, as well as the domains beyond the black hole horizon with $r<-r_c$. This system is entirely covered by the black hole Kruskal coordinates \eqref{eq:bh-krusk}.
\end{itemize}

With timelike spherically symmetric boundaries at $r=r_0$ in both static patches, the bulk equations \eqref{eq:classic-equation-for-metric}, \eqref{eq:classic-equation-for-dilaton} are unchanged and the local solution remains two-dimensional de Sitter with the same dilaton profile. The boundaries only affect the variational problem and the global data. In the case of Dirichlet boundary conditions, which we adopt, the induced metric $ds_b^2=-f(r_0)\,dt^2$ and the dilaton value $\Phi_b=\Phi(r_0)$ (that is defined by $\phi_r$ and $r_0$) are fixed on each boundary.

The setup with symmetric boundaries in full reduction model generalizes the half reduction model, in which two boundaries are located at $r=0$ \cite{Svesko:2022txo}, that corresponds to the cosmological system of the full reduction model with $r_0=0$. This allows us to consider both models uniformly and to investigate the dependence of the entanglement entropy of conformal fields on the location of the boundary.

It is convenient to introduce timelike and spacelike Kruskal coordinates adapted to the cosmological and black hole cases
\bea\label{eq:time-spacelike-Kruskal}
\tilde{T} =  \frac{\tilde{U}+\tilde{V}}{2}, \quad T =  \frac{U+V}{2}, \qquad \tilde{X}  = \frac{\tilde{V}-\tilde{U}}{2}, \quad X  = \frac{V-U}{2}.
\eea
Accordingly, for the cosmological case the Kruskal coordinates are given by
\bea\label{eq:timespacelike-to-dS-Lor}
\tilde{X} = \pm \frac{e^{-\kappa_c r_{*} (r)}}{\kappa_c} \cosh \kappa_c t, \qquad \tilde{T} = \mp \frac{e^{-\kappa_c r_{*} (r)}}{\kappa_c} \sinh \kappa_c t,
\eea
whereas for the black hole case they take the form
\bea\label{eq:timespacelike-to-dS-Lor-BH}
X = \pm \frac{e^{\kappa_c r_{*} (r)}}{\kappa_c} \cosh \kappa_c t, \qquad T = \pm \frac{e^{\kappa_c r_{*} (r)}}{\kappa_c} \sinh \kappa_c t.
\eea
In \eqref{eq:timespacelike-to-dS-Lor} and \eqref{eq:timespacelike-to-dS-Lor-BH}, the choice of sign corresponds to the right/left (upper/lower sign) static patches, where ``right" and ``left" are defined with respect to the origin of the corresponding Kruskal coordinate system — cosmological in the cosmological system and black hole in the black hole system.

To calculate the entanglement entropy using a replica trick that will be discussed later, it is convenient to consider the Euclidean theory. We  Wick rotate  Kruskal times $\tilde{T} = -i {\cal \tilde{T}}$, $T = -i {\cal T}$ and this also defines the Euclidean static time $\tau = i t$  periodic with a period of $2 \pi /\kappa_c$. The relation between the Euclidean Kruskal times and the Euclidean static-patch coordinates \((t,\tau)\) is given by
\bea
{\cal{\tilde{T}}} = \mp \frac{e^{-\kappa_c r_{*} (r)}}{\kappa_c} \sin \kappa_c \tau, \qquad
{\cal{T}} = \pm \frac{e^{\kappa_c r_{*} (r)}}{\kappa_c} \sin \kappa_c \tau.
\eea
In view of the relations
\bea\label{eq:euclidean-constant-surfaces}
\tilde{X}^2+{\cal \tilde{T}}^2 = \frac{e^{-2\kappa_c r_{*} (r)}}{\kappa^2_c}, \,\, \tan \kappa_c \tau = - \frac{{\cal \tilde{T}}}{\tilde{X}}, \quad X^2+{\cal T}^2 = \frac{e^{2\kappa_c r_{*} (r)}}{\kappa^2_c}, \,\, \tan \kappa_c \tau =  \frac{{\cal \tilde{T}}}{\tilde{X}},
\eea
the curves \(r=\mathrm{const}\) in the \(({\cal \tilde{T}},\tilde{X})\) and \(({\cal T}, X)\) planes are circles centered at the origin, while the curves \(\tau=\mathrm{const}\) are straight lines passing through the origin. The origin of the plane of cosmological Euclidean coordinates \(({\cal \tilde{T}},\tilde{X})\) corresponds to \(r=r_{c}\), since \(r_{*}(r)\to\infty\) as \(r\to r_{c}\) and the radius of the circle shrinks to zero. Similarly, in the black hole coordinates \((X,{\cal T})\), the origin corresponds to \(r=-r_{c}\).

In the cosmological and black hole cases, the radius of the circle corresponding to the position of the spherically symmetric boundaries \(r=r_{0}\) is  
\be
\tilde{L} = \frac{e^{-\kappa_{c} r_{*}(r_{0})}}{\kappa_{c}}, 
\qquad 
L = \frac{e^{\kappa_{c} r_{*}(r_{0})}}{\kappa_{c}}.
\ee
We introduce a complex coordinates
\be
w = X + i {\cal{T}}, \qquad \tilde{w} = \tilde{X}+i \cal{\tilde{T}}.
\ee
In the cosmological case, the Euclidean \(({\cal \tilde{T}},\tilde{X})\) plane contains the interior of a disk of radius \(\tilde{L}\), corresponding to the region \(r\geq r_{0}\). In complex coordinate, this can be written as $\tilde{w} \bar{\tilde{w}} \leq \tilde{L}^2$. Similarly, the black hole system in the Euclidean \(({\cal T},X)\) plane is described by the interior of a disk of radius \(L\), corresponding to the region \(r\leq r_{0}\), so $w \bar{w} \leq L^2$. Moreover, since the cosmological and black hole Kruskal coordinates within the same static patch are related by an inversion
\be
\tilde{U} =  -1/\kappa^2_c U, \qquad \tilde{V} =  -1/\kappa^2_c V,
\ee
in the same fixed Kruskal coordinates, for instance in the black hole chart $(U,V)$, the black hole system corresponds to the interior of a disk $w \bar{w} \leq L^2$, while the cosmological system corresponds to its exterior, $w \bar{w} \geq L^2$.

A very similar Euclidean geometric structure arises in the case of spherically symmetric boundaries in the two-sided Schwarzschild spacetime \cite{Ageev:2023hxe}.

\subsection{Semiclassical corrections}

We now include conformal matter in order to obtain semiclassical solutions.  We add two-dimensional boundary conformal field theory (BCFT) with central charge $c$, minimally coupled to action of the JT gravity. It means that equations of motion \eqref{eq:classic-equation-for-metric} changes as
\be\label{eq:semiclassical-equation}
g_{\mu\nu}\nabla^2 \phi -\nabla_{\mu}\nabla_{\nu} \phi+g_{\mu\nu}\Lambda \phi = \langle T_{\mu \nu} \rangle^{b.c.}_{g},
\ee
where $\langle T_{\mu \nu} \rangle^{b.c.}_{g}$ is the expectation value of the stress-energy tensor of BCFT with respect to some quantum state and boundary conditions. In this case, equation \eqref{eq:classic-equation-for-dilaton} retains its semiclassical form, so the backreaction affects only the dilaton, while the metric remains unchanged.

To determine the right-hand side of \eqref{eq:semiclassical-equation} we must specify the CFT state.
We impose conformally invariant boundary conditions at these boundaries in the BCFT (the details are given in the next subsection). The boundaries split spacetime into two decoupled domains, the cosmological and black hole systems, so we take a factorized global state
\be
\,|\Psi_{\text{full}}\rangle \;=\; |\mathrm{BD}^{\text{(bdy)}}\rangle_{\!\text{cosm}} \,\otimes\, |\mathrm{BD}^{\text{(bdy)}}\rangle_{\!\text{bh}}\,.
\ee
Each factor is the Bunch–Davies (BD) vacuum defined in the corresponding Kruskal coordinates. The boundaries correlate left- and right-moving modes within the same system by reflection, precisely as in the setup of Appendix A of \cite{Franken:2023ugu}, which analyzes two reflecting boundaries with the BD vacuum in the half reduction model.

In \cite{Batra:2024qju} the expectation value $\langle T_{\mu \nu} \rangle^{b.c.}_{g}$ was obtained in the vacuum state under conformally invariant boundary conditions on the timelike boundary, with the corresponding backreaction for a general BCFT fixed by the Polyakov action. The semiclassical corrections add a constant shift to the dilaton \cite{Batra:2024qju}, which now takes the form\footnote{In \cite{Batra:2024qju} it is claimed that the constant $\phi_r$ can be different from that in the classical solution \eqref{eq:static-sol}. For simplicity we choose the same $\phi_r$.}
\be\label{eq:static-sol-sem}
ds^2 = - f(r)\,dt^{2} + \frac{dr^2}{f(r)}, 
\qquad f(r) = 1 - \frac{r^2}{r_c^2}, 
\qquad \Phi(r) = \phi_0+\phi_r \frac{r}{r_c} + \frac{c G_2}{3}.
\ee


\subsection{Entanglement entropy}

To compute the entanglement entropy of conformal matter in JT gravity, we employ the island formula. The island contribution can be taken into account via the generalized entropy functional defined as
\be
    S_\gen[I, R] = \frac{\Phi(\partial I)}{4 G_2} + S_\m(R \cup I),
    \label{eq:gen_functional}
\ee
where $\partial I$ denotes the boundary of the entanglement island, and $S_\m$ denotes the entanglement entropy of conformal matter on a fixed background. One should extremize this functional over all possible island configurations
\be\label{eq:extremization-procedure}
    S\extgen[I, R] = \underset{\partial I}{\operatorname{ext}}\,\Big\{S_\gen[I, R]\Big\},
\ee
and then choose the minimal one
\be\label{eq:island-formula}
    S(R) = \underset{\partial I}{\text{min}}\,\Big\{S\extgen[I, R]\Big\}.
\ee

In the following, we discuss the evaluation of $S_\m$ within the framework of BCFT$_2$. The discussion in this subsection closely follows the corresponding subsection of \cite{Ageev:2023hxe}, so for further details we refer the reader to that work.

We begin with a brief overview of two-dimensional Euclidean boundary conformal field theory (BCFT) and the computation of entanglement entropy in this setting. The simplest example of a 2d BCFT is the theory formulated on the upper half-plane (UHP), equivalently described as evolution along the boundary line. The coordinate~$x_1$ corresponds to the Euclidean time, while $x_2$ is the spatial coordinate
\be
ds^2 = dx^2_1+dx^2_2 = dz d\bar{z}, \quad z = x_1 + i x_2, \quad x_1 \in (-\infty, \infty), \quad x_2 \geq 0,
\ee
We restrict attention to a region $R$ defined as a union of intervals
\be\label{eq:region}
R = [z_{a_1}, z_{b_1}] \cup \ldots \cup [z_{a_n}, z_{b_n}],
\ee
and its entanglement entropy
\be
S_\m(R) = -\Tr (\rho_R \log \rho_R),
\ee
where $\rho_R$ is reduced density matrix for region $R$. Entanglement entropy could be obtained by the replica trick \cite{Calabrese:2004eu, Calabrese:2009qy} as
\be\label{eq:entropy-from-replica-trick}
S_\m (R) = - \lim_{n \to 1} \partial_n \left(\Tr \rho^n_R\right).
\ee 
The trace in \eqref{eq:entropy-from-replica-trick} is expressed as a correlation function of twist operators inserted at the bulk endpoints of intervals of region $R$ that do not belong to the boundary $x_2=0$
\be\label{eq:correlator}
\Tr \rho^n_R = \langle \phi (z_{a_1}, \bar{z}_{a_1}) \phi (z_{b_1}, \bar{z}_{b_1}) \ldots \phi (z_{a_n}, \bar{z}_{a_n}) \phi (z_{b_n}, \bar{z}_{b_n})  \rangle_{\text{UHP}}.
\ee
If the endpoint of the interval is on the boundary, for example $R = [0, z_1]$, then the trace corresponding to such interval is given by  $\Tr \rho^n_R = \langle \phi (z_1, \bar{z}_1) \rangle_{\text{UHP}}$. The twist operators are primary operators with conformal dimensions $h_n=\bar{h}_n = c/24 (n-1/n)$, where $c$ is the central charge \cite{Calabrese:2009qy}.
\\

We consider the particular BCFT$_2$ given by $c$ copies of two-dimensional free massless Dirac fermions. The Dirac field is given by the doublet of the left and right moving components
\be
\psi (x_1, x_2) = \bigg(
\begin{array}{c}\psi_1(x_1, x_2) \\ \psi_2(x_1, x_2) \\ \end{array}
\bigg) = \bigg(
\begin{array}{c}\psi_1(z) \\ \psi_2(\bar{z}) \\ \end{array}
\bigg).
\ee
We choose the boundary conditions  preserving the conformal invariance of the theory and corresponding to the vanishing of the energy and momentum flow through the boundary \cite{Cardy:1984bb, Cardy:1986gw, Cardy:1989ir}. There are two explicit boundary conditions that satisfy this requirement for massless free Dirac fermions on UHP \cite{Mintchev:2020uom, Rottoli:2022plr}:
\be\label{eq:BC-vector-phase}
\psi_1 (x_1, 0) = e^{i \alpha_v} \psi_2 (x_1, 0), \qquad \alpha_v \in [0, 2\pi), \, \, x_1 \in (-\infty, \infty),
\ee
or
\be\label{eq:BC-axial-phase}
\psi_1 (x_1, 0) = e^{-i \alpha_a} \psi^{*}_2 (x_1, 0), \qquad \alpha_a \in [0, 2\pi), \, \, x_1 \in (-\infty, \infty).
\ee
These boundary conditions correspond to the transition of the left moving component to the right one, which can be interpreted as perfectly reflecting boundary conditions. The entanglement entropy of a region \eqref{eq:region} is given by \cite{Mintchev:2020uom, Rottoli:2022plr} 
\be
\begin{aligned}
S (R) = & \frac{1}{3} \sum^{n}_{i,j=1}\log |z_{a_i} - z_{b_j}| -\frac{1}{3}  \sum^{n}_{i<j} \log  |z_{a_i} - z_{a_j}| |z_{b_i} - z_{b_j}|-n \log \varepsilon  \\ + &\frac{1}{6} \sum^{n}_{i,j=1}\log |z_{a_i}-\bar{z}_{a_j}| |z_{b_i}-\bar{z}_{b_j}| -\frac{1}{6} \sum^{n}_{i,j=1}\log |z_{a_i}-\bar{z}_{b_j}| |z_{b_i}-\bar{z}_{a_j}|,
    \end{aligned}
    \label{eq:entropy-for-Dirac-fermions-UHP}
\ee
where $\varepsilon$ is UV cutoff. The entropy \eqref{eq:entropy-for-Dirac-fermions-UHP} does not depend on the choice of the boundary condition \eqref{eq:BC-vector-phase}  or \eqref{eq:BC-axial-phase} and on the phase $\alpha_{v, a}$ \cite{Mintchev:2020uom, Rottoli:2022plr}.
\\

If the state (corresponding to some geometry $\Omega$) of BCFT$_2$ is related with the UHP, then  under conformal transformation 
\be\label{eq:conf-transf-general}
z: \Omega \to \text{UHP}, \quad  z = z(w), \quad \bar{z} = \bar{z} (\bar{w}).
\ee
 correlation function of twist operators  transforms as
\be
\begin{aligned}
\langle \phi (w_1, \bar{w}_1) \ldots \phi (w_m, \bar{w}_m) \rangle_{\Omega} =  \prod\limits_{j = 1}^m & \left( \frac{dz}{dw}\right)^{h_n}\Big|_{w=w_j} \left( \frac{d\bar{z}}{d\bar{w}}\right)^{\bar{h}_n}\Big|_{\bar{w}=\bar{w}_j}   \\ \times & \langle \phi (z_1, \bar{z}_1) \ldots \phi (z_m, \bar{z}_m) \rangle_{\text{UHP}}. 
    \end{aligned}
   \label{eq:conf}
\ee
In our case, the required conformal map is from the interior of the disk $w \bar{w} \leq L^2$ to the upper half-plane
\be
z = i \frac{L+w}{L-w}.
\ee
Entanglement entropies in flat $ds^2 = dw d\bar{w}$ and curved $ds^2 = e^{2 \rho (w, \bar{w}) } dw d\bar{w}$ two-dimensional spacetimes are related by Weyl transformation $ds^2 \to e^{2\rho} ds^2$
\be \label{eq:weyl}
S_\m\Big|_{ds^2 = e^{2 \rho (w, \bar{w}) } dw d\bar{w}}= S_\m\Big|_{ds^2 =  dw d\bar{w}} + \frac{c}{6} \sum_{i=1}^m \log e^{\rho (w_i, \bar{w}_i) }.
\ee
In our case of two-dimensional de Sitter spacetime, the Weyl factor is
\be
e^{2 \rho (w, \bar{w}) } = \frac{4}{(1+\kappa^2_c w \bar{w})^2}.
\ee


\section{Black hole system}\label{sec:bh}

\begin{figure}[h!]\centering
    \includegraphics[width=0.8\textwidth]{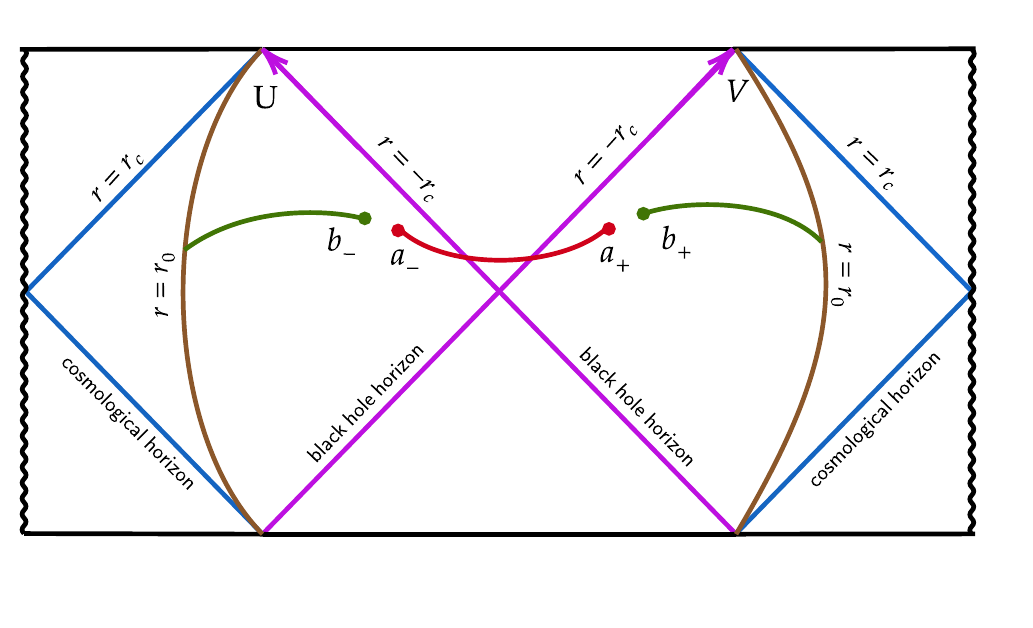}
	\caption{Symmetric entanglement region and island for a black hole system in the presence of timelike boundaries in static patches.}
    \label{fig:full-reduction-penrose-bh}
\end{figure}

We now consider the region \(R_{\text{BH}}\) in the full reduction model of JT gravity in de Sitter space, obtained after introducing symmetric reflecting boundaries at \(r = r_0\) in the static patches, with \(r_0\) chosen close to the cosmological horizon \(r = r_c\). This construction isolates the portion of the spacetime containing the black hole horizon at \(r = -r_c\). The region \(R_{\text{BH}}\) is defined as the union of two intervals extending from bulk points \(b_\pm\) to the reflecting boundary in the corresponding left/right wedges, see Fig.\ref{fig:full-reduction-penrose-bh}. In Kruskal coordinates these points are chosen as  
\begin{equation}
b_+ = (U_b,\, V_b), \qquad b_- = (V_b,\, U_b),
\end{equation}
so that they are symmetric under interchange of \(U\) and \(V\). In static coordinates this corresponds to  
\begin{equation}
b_+ = (r_b,\, t_b), \qquad b_- = (r_b,\, -t_b).
\end{equation}
The radial position \(r_b\) is taken sufficiently close to the black–hole horizon at \(r = -r_c\), which means that \(R_{\text{BH}}\) covers nearly the entire static patches inside the domain bounded by the reflecting surfaces. 

We first analyze the entanglement entropy of \(R_{\text{BH}}\) without including an island contribution. We then proceed to study the case with an island configuration. In both cases, the computation follows the method described in the Setup, adapted to this symmetric geometry.

\subsection{Without island}
\begin{figure}[h!]\centering
    \includegraphics[width=0.4825\textwidth]{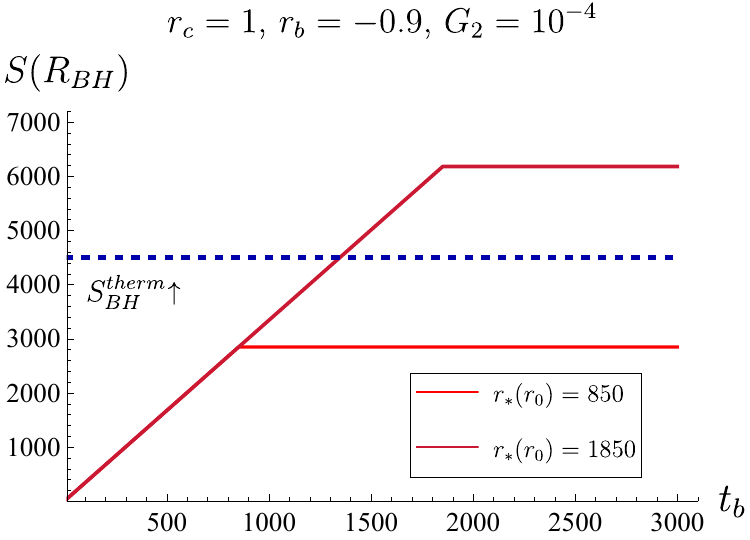} \quad \includegraphics[width=0.4825\textwidth]{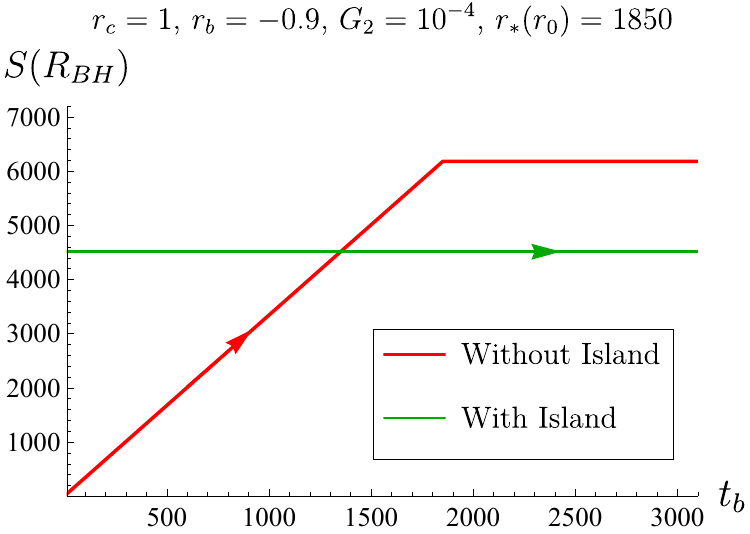}
	\caption{\textbf{Left}: time evolution of the entanglement entropy of the region \(R_{\text{BH}}\) without including an island configuration, shown for different boundary positions (larger \(r_*(r_0)\) corresponds to the boundary placed closer to the cosmological horizon). For sufficiently large \(r_*(r_0)\) the entanglement entropy exceeds the thermodynamic entropy of the black hole horizon, \(S^{\text{therm}}_{\text{BH}}\). \textbf{Right}: time evolution of the entanglement entropy with the island configuration included (green curve). The island dominates only at late times, and the corresponding entropy is approximately equal to the thermodynamic entropy of the black hole horizon. Parameters used: \(r_c = 1\), \(r_b = -0.9\), \(c = 10\), \(G_2 = 10^{-4}\), \(\phi_0 = 1\), \(\phi_r = 1\).}
    \label{fig:entropy-bh}
\end{figure}

In the absence of an island configuration, the entanglement entropy of $R_{\text{BH}}$ is given by
\begin{equation}\label{eq:no-island-bh-kruskal}
S(R_{\text{BH}}) = \frac{c}{3}\log\!\left[\frac{2\,(V_b-U_b)}{(1-\kappa_c^2 U_b V_b)\varepsilon}\right]
+\frac{c}{6}\log\!\left[\frac{(L^2+U_b V_b)^2}{(L^2+U_b^2)(L^2+V_b^2)}\right].
\end{equation}
It is convenient to rewrite the entropy \eqref{eq:no-island-bh-kruskal}, expressed in Kruskal coordinates \((U_b, V_b)\), in terms of the static coordinates \((t_b, r_b)\)
\begin{equation}\label{eq:bh-with-no-island-static}
\begin{aligned}
S(R_{\text{BH}}) &= \frac{c}{3}\log\!\Bigg[
    \frac{2\cosh \kappa_c t_b}
         {\kappa_c \,\varepsilon \cosh \kappa_c r_*(r_b)}
    \Bigg]+\frac{c}{6}\log\!\Bigg[
    \frac{2\sinh^2 \kappa_c (r_*(r_0)-r_*(r_b))}
         {\cosh 2\kappa_c (r_*(r_0)-r_*(r_b)) + \cosh 2\kappa_c t_b}
    \Bigg].
\end{aligned}
\end{equation}
We summarize below the key aspects of the entanglement entropy without an island for the region $R_{\text{BH}}$, which is also shown on Fig.\ref{fig:entropy-bh} (left):
\begin{itemize}
    \item At early times \(t_b \ll t_b^{h}\), with \(t^h_b \equiv r_*(r_0) - r_*(r_b) > 0\) (equivalently, \(V_b \ll L\) in Kruskal coordinates), the entropy \eqref{eq:bh-with-no-island-static} increases monotonically in time, becoming linear when \(\kappa_c t_b \gg 1\)
    \begin{equation}\label{eq:bh-with-no-island-static-early}
\begin{aligned}
S(R_{\text{BH}}) &\simeq \frac{c\kappa_c t_b}{3} + \frac{c}{3}\,\log\!\Bigg[
    \frac{1}
         {\kappa_c \varepsilon \cosh \kappa_c r_*(r_b)}
    \Bigg].
\end{aligned}
\end{equation}
    
    \item In the late–time regime \(t_b \gg t_b^{h}\) (equivalently, \(V_b \gg L\)), the entropy \eqref{eq:bh-with-no-island-static} saturates to the value
\begin{equation}\label{eq:sat-entr}
S_{\text{sat}}(R_{\text{BH}}) \;=\; \frac{c}{3}\,\log\!\left[
\frac{2 \sinh\!\kappa_c\,(r_*(r_0)-r_*(r_b))}
{\kappa_c\,\varepsilon \cosh \kappa_c r_*(r_b)}
\right].
\end{equation}
In the regime \(\kappa_c(r_*(r_0)-r_*(r_b)) \gg 1\) that we focus on in this analysis (with \(r_b\) close to \(r = -r_c\) and \(r_0\) close to \(r = r_c\)), the saturation value of the entropy simplifies to
\begin{equation}\label{eq:big-r0}
S_{\text{sat}}(R_{\text{BH}}) \;\simeq\; \frac{c}{3}\,\log\!\left[
\frac{1}
{\kappa_c\,\varepsilon \cosh \kappa_c r_*(r_b)}
\right] + \frac{c \kappa_c\,(r_*(r_0)-r_*(r_b))}{3}.
\end{equation}

\item In the limit \(r_*(r_0) \to \infty\) (equivalently, \(r_0 \to r_c\)), for which \(t_b^{h} \to \infty\), the second term in \eqref{eq:bh-with-no-island-static} vanishes for any finite \(t_b\). The expression therefore reduces to its first term, representing a regime of monotonic growth. Equivalently, this limit lies entirely within the early–time phase \eqref{eq:bh-with-no-island-static-early}.

\item Since $r_*(r_0)$ can be made arbitrarily large in the limit $r_0 \to r_c$, the saturation value of the entropy \eqref{eq:big-r0} can likewise grow without bound. For sufficiently large $r_*(r_0)$, this saturation entropy exceeds the thermodynamic entropy of the black hole horizon, $S (R_{\text{BH}}) > S^{\text{therm}}_{\text{BH}}$, where
\be\label{eq:therm-entropy-of-BH}
S^{\text{therm}}_{\text{BH}} = \frac{\Phi(-r_c)}{2 G_2} \qquad \Rightarrow \qquad S^{\text{therm}}_{\text{BH}} = \frac{\phi_0 - \phi_r}{2G_2} + \frac{2 c G_2}{3}.
\ee

\end{itemize}

Thus, we find that the entanglement entropy of the region \(R_{\text{BH}}\) can exceed the thermodynamic entropy of the horizon \(S^{\text{therm}}_{\text{BH}}\). This situation is directly analogous to the information paradox in a two–sided black hole, in particular to the setup considered in \cite{Ageev:2023hxe}, where a two–sided Schwarzschild black hole with symmetric perfectly reflecting boundaries was analyzed. If the composite “black hole + radiation” system is closed and its global state is pure, unitarity requires the entanglement entropies of the two subsystems to coincide and to be bounded above by the smaller of their thermodynamic entropies. A violation of this bound thus signals non–unitary evolution.

In the present setup — the full reduction model of JT gravity in de Sitter space — we examine the black hole region defined by the reflecting boundaries and find that its entanglement entropy can exceed the thermodynamic entropy of the black hole horizon. We interpret this excess as the analogue of the information paradox, formulated in terms of the time evolution of the entanglement entropy.

\subsection{With island}

In the preceding subsection we have shown that, for \(r_0\) sufficiently close to the cosmological horizon, the entanglement entropy of \(R_{\mathrm{BH}}\) in the absence of an island can exceed the thermodynamic entropy of the black hole horizon. In this subsection we investigate whether a nontrivial island configuration exists for \(R_{\mathrm{BH}}\), and, if so, whether its contribtution can eliminate this excess.

The symmetry of the reflecting boundaries in the left and right static patches, together with the corresponding symmetry of the endpoints \(b_{\pm}\) that define the region \(R_{\mathrm{BH}}\), implies that the island considered should inherit the \(U\!\leftrightarrow\!V\) exchange symmetry. We therefore restrict attention to a simply connected island \(I_{\text{BH}}=[a_{-},a_{+}]\), which is contained in the causal complement of \(R_{\mathrm{BH}}\) and possesses the above symmetry, see Fig.\ref{fig:full-reduction-penrose-bh}.

In Kruskal coordinates the island endpoints are chosen as
\[
a_{+}=(U_{a},\,V_{a}),\qquad a_{-}=(V_{a},\,U_{a}),
\]
so that \(a_{-}\) and \(a_{+}\) are mapped into one another by \(U_a\!\leftrightarrow\!V_a\). Equivalently, in static coordinates the endpoints take the form
\[
a_{+}=(r_{a},\,t_{a}),\qquad a_{-}=(r_{a},\,-t_{a}),
\]
with the radial coordinate constrained by \(r_{a}<r_{b}\) to ensure that the island remains within the causal complement of the exterior region \(R_{\mathrm{BH}}\). The specific coordinates of the island endpoints, and their dependence on the parameters of the full-reduction JT model with symmetric reflecting walls, will be determined by the extremization of the generalized entropy
\begin{equation}\label{eq:Sgen_simple}
S_{\mathrm{gen}}(R_{\text{BH}})\;=\;S_{\mathrm{area}}(\partial I_{\text{BH}})\;+\;S_{\mathrm{matter}}(I_{\text{BH}}\cup R_{\text{BH}}),
\end{equation}
where the area term is given by
\begin{equation}\label{eq:S_area}
 S_{\mathrm{area}}(\partial I_{\text{BH}}) \;=\;  \frac{\phi_0}{2G_2}
-\frac{\phi_r}{2G_2}\,\frac{1+\kappa_c^2 U_a V_a}{1-\kappa_c^2 U_a V_a} + \frac{2c G_2}{3}, 
\end{equation}
while the matter contribution takes the form
\begin{flalign}\label{eq:Smatter_leftmost}
& S_{\mathrm{matter}}(I_{\text{BH}}\cup R_{\text{BH}})
= \frac{c}{6}\log\!\Bigg[
\frac{16\,(U_a-U_b)^2\,(U_a-V_a)^2\,(U_b-V_b)^2\,(V_a-V_b)^2}
{\varepsilon^4 (U_b-V_a)^2\,(U_a-V_b)^2\,(1-\kappa_c^2 U_a V_a)^2\,(1-\kappa_c^2 U_b V_b)^2}
\Bigg] & \\[6pt]
& +\;\frac{c}{6}\log\!\Bigg[
\frac{%
\bigl(L^2+U_aU_b\bigr)^{2}\,
\bigl(L^2+U_aV_a\bigr)^{2}\,
\bigl(L^2+U_bV_b\bigr)^{2}\,
\bigl(L^2+V_aV_b\bigr)^{2}
}
{%
\bigl(L^2+U_a^{2}\bigr)\,
\bigl(L^2+U_b^{2}\bigr)\,
\bigl(L^2+U_bV_a\bigr)^{2}\,
\bigl(L^2+V_a^{2}\bigr)\,
\bigl(L^2+U_aV_b\bigr)^{2}\,
\bigl(L^2+V_b^{2}\bigr)
}
\Bigg]. \nn &
\end{flalign}

The extremization of the generalized entropy with respect to the island endpoints leads to the system
\begin{equation}
\begin{cases}
\partial_{U_a} S_{\mathrm{gen}}\!\left(U_a, V_a, U_b, V_b, L, c, G_2, \phi_r, \kappa_c\right) = 0, \\[4pt]
\partial_{V_a} S_{\mathrm{gen}}\!\left(U_a, V_a, U_b, V_b, L, c, G_2, \phi_r, \kappa_c\right) = 0,
\end{cases}
\label{eq:extremization_system}
\end{equation}
which will be analyzed in the regime \(c\,G_2 / \phi_r \ll 1\), where the semiclassical approximation remains valid. Analytical and numerical investigations of this extremization problem lead to the following solutions
\begin{equation}\label{eq:sol-for-isl}
U_a=U_b+\frac{cG_2}{3\phi_r}\frac{(1-\kappa_c^2U_bV_b)^2}{\kappa_c^2V_b}
+ O\!\left(\frac{c^2 G_2^2}{\phi_r^2}\right), \quad
V_a=V_b+\frac{cG_2}{3\phi_r}\frac{(1-\kappa_c^2U_bV_b)^2}{\kappa_c^2U_b}
+ O\!\left(\frac{c^2 G_2^2}{\phi_r^2}\right).
\end{equation}
which, when expressed in static coordinates, take the form
\begin{equation}\label{eq:stat-sol-bh}
r_a = r_b - \frac{4cG_2}{3\phi_r} r_c + \mathit{O}\!\left(\frac{c^2G_2^2}{\phi_r^2}\right),\; \quad
t_a = t_b + \mathit{O}\!\left(\frac{c^2G_2^2}{\phi_r^2}\right).
\end{equation}
The endpoints \(a_{\pm}\) are positioned very close to their counterparts \(b_{\pm}\), with \(a_{+}\) and \(a_{-}\) situated in the right and left static patches, respectively. The island is indeed contained within the causal complement of the region \(R_{\mathrm{BH}}\) due to \(r_a < r_b, \, t_a \simeq t_b\).

Substituting the explicit solutions~\eqref{eq:sol-for-isl} into the generalized entropy~\eqref{eq:Sgen_simple} yields the expression for the entropy with the inclusion of the island configuration:
\begin{equation}\label{eq:entropy-with-island}
\begin{aligned}
S_{\text{Island}} (R_{\text{BH}}) &= \frac{\phi_0}{2G_2}
-\frac{\phi_r}{2G_2}\,\frac{1+\kappa_c^2 U_b V_b}{1-\kappa_c^2 U_b V_b}
+ \frac{c}{3} \log\!\left[
\frac{4}{9\,\kappa_c^{4}\,\varepsilon^{2}\,|U_b| V_b}
\left(\frac{c G_2}{\phi_r^{2}}\right)^{\!2}
\left(1-\kappa_c^{2} U_b V_b \right)^{\!2}
\right] \\
&\hspace{4cm}
+ \frac{2c G_2}{3}
-\frac{2c}{3}
+ \frac{2c^{2} G_2}{9\phi_r}\,\left(1-\kappa_c^{2} U_b V_b\right)
+ \mathit{O}\!\left(\frac{c^{2} G_2^{2}}{\phi_r^{2}}\right).
\end{aligned}
\end{equation}
From \eqref{eq:entropy-with-island} we find that, when the points \(b_{\pm}\) are located sufficiently close to the horizons (so that \(\kappa_c^2 U_b V_b \approx 0\)), the leading contribution to the entropy with island coincides with the thermodynamic entropy of the black hole horizon~\eqref{eq:therm-entropy-of-BH}. The remaining terms in \eqref{eq:entropy-with-island} are logarithmic and power-law corrections. In the leading orders the island entropy is independent of the time coordinate \(t_b\), since the dependence on the parameters of \(b_{\pm}\) enters only through the combination \([U_b V_b](r_b)\), which does not vary with~\(t_b\), see Fig.\ref{fig:entropy-bh} (right).

Within the island formula, the entanglement entropy is defined as the minimal value of the generalized-entropy functional evaluated on all extrema. When the no-island entropy reaches the thermodynamic entropy of the horizon (the Page time), the extremum that includes a nontrivial island becomes the global minimum and therefore determines the entanglement entropy thereafter. From that moment the entanglement entropy of \(R_{\mathrm{BH}}\) ceases to increase and saturates at the island value. Because the island contribution prevents the entanglement entropy from ever exceeding the horizon thermodynamic entropy within the cavity-regulated JT model considered here, the apparent violation of unitarity exhibited by the no-island calculation is avoided.

\section{Cosmological system}\label{sec:cosm}

\begin{figure}[h!]\centering
    \includegraphics[width=0.8\textwidth]{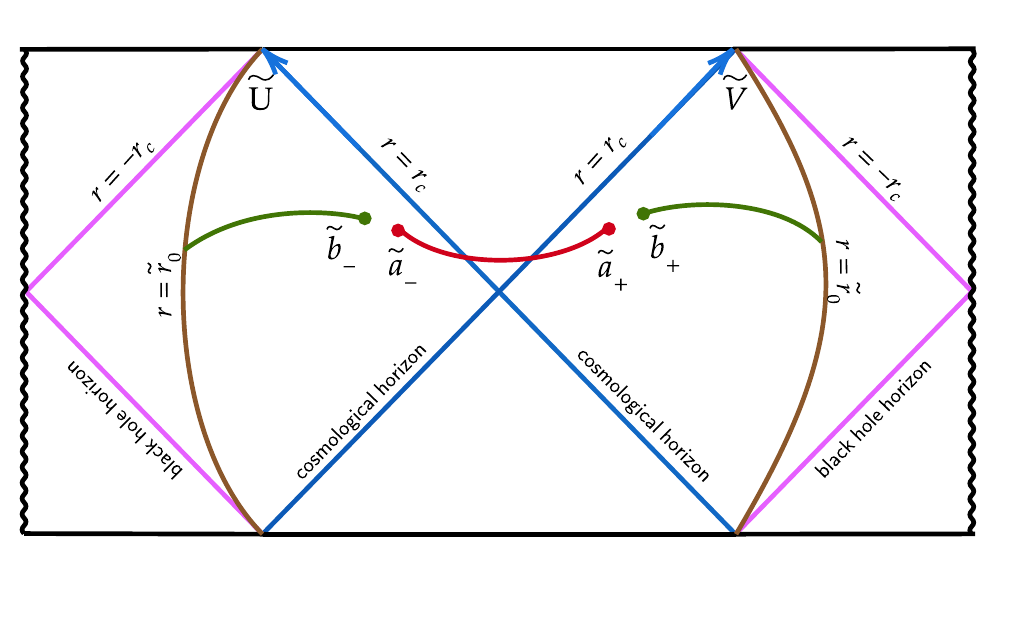}
	\caption{Symmetric entanglement region and island for a cosmological system in the presence of timelike boundaries in static patches.}
    \label{fig:full-reduction-penrose-cosm}
\end{figure}

We now turn to the entanglement entropy of the cosmological system of the full reduction JT model, the domain that contains only the cosmological horizon at $r = r_c$, with symmetric boundaries at $r = \tilde{r}_0$ in both the static patches. We consider a region \(R_{\mathrm{cosm}}\) of the same type as the black hole region analysed in the previous chapter: it extends from two bulk endpoints \(\tilde{b}_{\pm}\) inside the static patches, with \(\tilde{b}_{+}\) in the right static patch and \(\tilde{b}_{-}\) in the left static patch, to the boundaries in the corresponding patches, see~Fig.\ref{fig:full-reduction-penrose-cosm}. In Kruskal coordinates the endpoints are chosen as
\[
\tilde{b}_+ = (\tilde{U}_b,\,\tilde{V}_b),\qquad \tilde{b}_- = (\tilde{V}_b,\,\tilde{U}_b),
\]
and in static coordinates as
\[
\tilde{b}_+ = (\tilde{r}_b,\,t_b),\qquad \tilde{b}_- = (\tilde{r}_b,\,-t_b).
\]

The main structural distinction from the black hole system setup is that the roles of the two horizons, bulk endpoints $\tilde{b}_{\pm}$ and boundaries are reversed: now  \(\tilde{b}_{\pm}\) are located near the cosmological horizon, so $\tilde{r}_b \approx r_c$, while the reflecting boundaries are placed near the black-hole horizon so that $\tilde{r}_0 \approx -r_c$. Because the two regions are constructed in close parallel, much of the entanglement-entropy computation carries over, though important differences remain.

\subsection{No island}

If no island is included, the entanglement entropy of \(R_{\text{cosm}}\) takes the form
\begin{equation}\label{eq:no-island-c-kruskal}
S(R_{\text{cosm}}) = \frac{c}{3}\log\!\left[\frac{2\,(\tilde{V}_b-\tilde{U}_b)}{(1-\kappa_c^2 \tilde{U}_b \tilde{V}_b)\varepsilon}\right]
+\frac{c}{6}\log\!\left[\frac{(\tilde{L}^2+\tilde{U}_b \tilde{V}_b)^2}{(\tilde{L}^2+\tilde{U}_b^2)(\tilde{L}^2+\tilde{V}_b^2)}\right],
\end{equation}
or in static coordinates
\begin{equation}\label{eq:c-with-no-island-static}
\begin{aligned}
S(R_{\text{cosm}}) &= \frac{c}{3}\log\!\Bigg[
    \frac{2\cosh \kappa_c t_b}
         {\kappa_c \,\varepsilon \cosh \kappa_c r_*(\tilde{r}_b)}
    \Bigg]+\frac{c}{6}\log\!\Bigg[
    \frac{2\sinh^2 \kappa_c (r_*(\tilde{r}_0)-r_*(\tilde{r}_b))}
         {\cosh 2\kappa_c (r_*(\tilde{r}_0)-r_*(\tilde{r}_b)) + \cosh 2\kappa_c t_b}
    \Bigg].
\end{aligned}
\end{equation}
Because the entropy \eqref{eq:c-with-no-island-static} is invariant under the replacement \(\tilde r_0 \to -\tilde r_0\), \(\tilde r_b \to -\tilde r_b\), the behavior of the no–island entropy in the cosmological system closely parallels that of the black hole system.
In the early–time regime \(t_b \ll t_b^{c}\), with  
\(
t_b^{c} \equiv r_*(\tilde{r}_b) - r_*(\tilde{r}_0) > 0\), (equivalently, $\tilde{U}_b \gg \tilde{L}$), the entropy grows monotonically, while in the late–time regime \(t_b \gg t_b^{c}\) the entropy approaches a saturation value, directly analogous to \eqref{eq:sat-entr} and \eqref{eq:big-r0}
\begin{equation}\label{eq:big-r0-cosm}
S_{\text{sat}}(R_{\text{cosm}}) \;\simeq\; \frac{c}{3}\,\log\!\left[
\frac{1}
{\kappa_c\,\varepsilon \cosh \kappa_c r_*(\tilde{r}_b)}
\right] + \frac{c \kappa_c\,(r_*(\tilde{r}_b)-r_*(\tilde{r}_0))}{3}.
\end{equation}
Thus, as in the black hole system, the no–island entropy in the cosmological case can become arbitrarily large at late times if the boundary \(\tilde r_0\) is placed sufficiently close to the black–hole horizon at \(r=-r_c\). In particular, the no–island entropy can exceed the thermodynamic entropy of the cosmological horizon,  
\be\label{eq:therm-entropy-of-c}
S^{\text{therm}}_{\text{cosm}} = \frac{\phi_0 + \phi_r}{2G_2} + \frac{2c G_2}{3}.
\ee
This implies a breakdown of unitarity in the evolution of the closed “cosmological horizon + radiation” system, which we interpret as the analogue of the information paradox in this setup.

\subsection{With island}
In the previous subsection we established that, under certain conditions, the no–island entropy of \(R_{\text{cosm}}\) can exceed the thermodynamic entropy of the cosmological horizon. The central question we examine here is whether an island configuration exists that keeps the entanglement entropy below this thermodynamic bound at all times, thereby restoring consistency with unitarity.

Applying the same reasoning as in the black hole system, we restrict to a simply connected, symmetric island located in the causal complement of the region \(R_{\text{cosm}}\). We therefore consider an island \(I_{\text{cosm}} = [\tilde{a}_{-},\tilde{a}_{+}]\) with endpoints chosen in Kruskal coordinates as  
\[
\tilde{a}_{+} = (\tilde{U}_{a}, \tilde{V}_{a}), \qquad \tilde{a}_{-} = (\tilde{V}_{a}, \tilde{U}_{a}),
\]
so that the pair is related by the exchange \(U_a \leftrightarrow V_a\), see Fig.\ref{fig:full-reduction-penrose-cosm}. In static coordinates the endpoints take the form  
\[
\tilde{a}_{+} = (\tilde{r}_{a}, \tilde{t}_{a}), \qquad \tilde{a}_{-} = (\tilde{r}_{a}, -\tilde{t}_{a}),
\]
with the condition \(\tilde{r}_{a} > \tilde{r}_{b}\), ensuring that the island lies entirely within the causal complement of the exterior region \(R_{\text{cosm}}\). To investigate whether such an island configuration exists, we now write down the generalized entropy
\begin{equation}\label{eq:Sgen_simple_cosm}
S_{\mathrm{gen}}(R_{\text{cosm}})\;=\;S_{\mathrm{area}}(\partial I_{\text{cosm}})\;+\;S_{\mathrm{matter}}(I_{\text{cosm}}\cup R_{\text{cosm}}),
\end{equation}
where the area term is given by
\begin{equation}\label{eq:S_area-cosm}
 S_{\mathrm{area}}(\partial I_{\text{cosm}}) \;=\;  \frac{\phi_0}{2G_2}
+\frac{\phi_r}{2G_2}\,\frac{1+\kappa_c^2 \tilde U_a \tilde V_a}{1-\kappa_c^2 \tilde U_a \tilde V_a} + \frac{2c G_2}{3}, 
\end{equation}
while the matter contribution takes the form
\begin{flalign}\label{eq:Smatter_leftmost}
& S_{\mathrm{matter}}(I_{\text{cosm}}\cup R_{\text{cosm}})
= \frac{c}{6}\log\!\Bigg[
\frac{16\,(\tilde U_a-\tilde U_b)^2\,(\tilde U_a-\tilde V_a)^2\,(\tilde U_b-\tilde V_b)^2\,(\tilde V_a-\tilde V_b)^2}
{\varepsilon^4 (\tilde U_b-\tilde V_a)^2\,(\tilde U_a-\tilde V_b)^2\,(1-\kappa_c^2 \tilde U_a \tilde V_a)^2\,(1-\kappa_c^2 \tilde U_b \tilde V_b)^2}
\Bigg] & \\[6pt]
& +\;\frac{c}{6}\log\!\Bigg[
\frac{%
\bigl(\tilde{L}^2+\tilde U_a\tilde U_b\bigr)^{2}\,
\bigl(\tilde{L}^2+\tilde U_a\tilde V_a\bigr)^{2}\,
\bigl(\tilde{L}^2+\tilde U_b\tilde V_b\bigr)^{2}\,
\bigl(\tilde{L}^2+\tilde V_a\tilde V_b\bigr)^{2}
}
{%
\bigl(\tilde{L}^2+\tilde U_a^{2}\bigr)\,
\bigl(\tilde{L}^2+\tilde U_b^{2}\bigr)\,
\bigl(\tilde{L}^2+\tilde U_b\tilde V_a\bigr)^{2}\,
\bigl(\tilde{L}^2+\tilde V_a^{2}\bigr)\,
\bigl(\tilde{L}^2+\tilde U_a\tilde V_b\bigr)^{2}\,
\bigl(\tilde{L}^2+\tilde V_b^{2}\bigr)
}
\Bigg]. \nn &
\end{flalign}
We note that the generalized entropy \eqref{eq:Sgen_simple_cosm} closely resembles its counterpart in the black hole system \eqref{eq:Sgen_simple}, differing only in the sign of the area term \eqref{eq:S_area-cosm}. This difference originates from the distinct functional relation between the dilaton and the Kruskal coordinates associated with the cosmological and black hole horizons, as discussed in the Setup.

A \textit{formal} solution of the extremization problem for the generalized entropy \eqref{eq:Sgen_simple_cosm} in the regime \(c\,G_2 / \phi_r \ll 1\) yields  
\begin{equation}\label{eq:sol-for-isl-cosm}
\tilde{U}_a=\tilde{U}_b-\frac{cG_2}{3\phi_r}\frac{(1-\kappa_c^2\tilde{U}_b\tilde{V}_b)^2}{\kappa_c^2\tilde{V}_b}
+ O\!\left(\frac{c^2 G_2^2}{\phi_r^2}\right), \quad
\tilde{V}_a=\tilde{V}_b-\frac{cG_2}{3\phi_r}\frac{(1-\kappa_c^2\tilde{U}_b\tilde{V}_b)^2}{\kappa_c^2\tilde{U}_b}
+ O\!\left(\frac{c^2 G_2^2}{\phi_r^2}\right).
\end{equation}
Relative to \eqref{eq:sol-for-isl}, the subleading contributions appear with opposite signs.  In static coordinates the solution takes the form  
\begin{equation}\label{eq:sol-stat-cosm}
\tilde{r}_a = \tilde{r}_b - \frac{4cG_2}{3\phi_r} r_c + \mathit{O}\!\left(\frac{c^2G_2^2}{\phi_r^2}\right),\; \quad
\tilde{t}_a = t_b - \mathit{O}\!\left(\frac{c^2G_2^2}{\phi_r^2}\right),
\end{equation}
where, which is of particular importance for what follows, the expression for the radial coordinate coincides with that in \eqref{eq:stat-sol-bh}.

However, this solution does not satisfy the conditions under which the generalized entropy \eqref{eq:Sgen_simple_cosm} was derived. Specifically, the formula \eqref{eq:sol-stat-cosm} yields \(\tilde{r}_a < \tilde{r}_b\), whereas the construction assumed that the island \(I_{\text{cosm}}\) is contained in the causal complement of the region \(R_{\text{cosm}}\), which requires \(\tilde{r}_a > \tilde{r}_b\). We therefore conclude that the solution \eqref{eq:sol-for-isl-cosm}, \eqref{eq:sol-stat-cosm} is unphysical and \textbf{must be excluded}.

The solution \eqref{eq:sol-for-isl-cosm}, \eqref{eq:sol-stat-cosm} can be viewed as inherited, in a certain sense, from the one found in the black hole system \eqref{eq:sol-for-isl}, \eqref{eq:stat-sol-bh}, since the generalized entropy has essentially the same structure in both cases. However, the differences between the black hole and cosmological systems imply that the corresponding solution in the latter does not satisfy the island conditions.

Since the exclusion of the solution \eqref{eq:sol-for-isl-cosm}, \eqref{eq:sol-stat-cosm} as unphysical may not be entirely clear — especially why an island boundary with \(\tilde r_a < \tilde r_b\) is not admissible — we now discuss this point in more detail. Let us proceed by contradiction and assume initially that the region \(R_{\text{cosm}}\) admits a simply connected, symmetric island \(\hat{I}_{\text{cosm}}\) with \(\tilde r_a < \tilde r_b\). In this case the generalized entropy takes the form  
\begin{equation}\label{eq:Sgen_simple_cosm_bad_choice}
S_{\mathrm{gen}}(R_{\text{cosm}})\;=\;S_{\mathrm{area}}(\partial \hat{I}_{\text{cosm}})\;+\;S_{\mathrm{matter}}(\hat{I}_{\text{cosm}}\cup R_{\text{cosm}}).
\end{equation}
Consider the matter contribution \(S_{\mathrm{matter}}(\hat{I}_{\text{cosm}}\cup R_{\text{cosm}})\). Since \(\tilde r_a < \tilde r_b\), the spacelike surface \(\hat{I}_{\text{cosm}}\cup R_{\text{cosm}}\) coincides with the entire Cauchy slice belonging to the cosmological system. Our framework assumes that the quantum fields on this slice are in a pure state, so in general  
\(
S_{\mathrm{matter}}(\hat{I}_{\text{cosm}}\cup R_{\text{cosm}})=0.
\)  
Alternatively, the argument can be phrased using the well-established fact that the entanglement entropy of a region is determined entirely by its causal domain \cite{Almheiri:2020cfm}. Varying the island endpoints \((\tilde r_a, \tilde t_a)\) (or equivalently \((\tilde U_a,\tilde V_a)\)) does not change \(S_{\mathrm{matter}}(\hat{I}_{\text{cosm}}\cup R_{\text{cosm}})\), since the causal domain remains unchanged. Indeed, modifying the coordinates leaves the underlying spacelike region itself unaltered. Consequently, this matter term gives zero when differentiated with respect to the island coordinates. While the area term, determined by the dilaton, is a monotonic function of \(\tilde r_a\), so its derivative therefore never vanishes.  

Thus, it follows that even if the island boundaries were hypothetically placed within the causal domain of \(R_{\text{cosm}}\), no extremal configuration would arise. Therefore, the excess of thermodynamic entropy of the cosmological horizon for some locations of the boundary cannot be removed using the island configuration.

\section{Conclusions and discussion}

We analyzed entanglement entropy of conformal matter in Jackiw–Teitelboim de Sitter with symmetric timelike reflecting boundaries that partition the geometry into a black hole system and a cosmological system. Using the island formula and BCFT methods, we found that in the black hole system an island dominates at late times and leads to saturation of the entanglement entropy at the horizon value, thereby preventing late-time growth and addressing the potential information paradox in this boundary-regulated setup. In contrast, in the cosmological system no island is found, and the entanglement entropy can be made arbitrarily large by placing the boundary sufficiently close to the black hole horizon. This signals a tension with unitarity in that sector and highlights a qualitative difference between the two systems defined by the same JT background. A similar result was obtained earlier in the half reduction model of JT gravity in de Sitter space, where the non-existence of islands was also shown \cite{Ageev:2023mzu}.

The black hole system reproduces the expected island mechanism of two-sided eternal configurations in a setting with positive cosmological constant. Unlike the ``blinking island’’ phenomenon observed for two-sided Schwarzschild with a reflecting wall \cite{Ageev:2023hxe}, the island here persists and ensures saturation for all late times. The differing behavior with respect to the blinking island effect is paralleled by a distinction in thermodynamic properties between the Schwarzschild setup with reflecting boundaries and its JT de Sitter counterpart. In the Schwarzschild case the heat capacity changes sign with the boundary distance — negative when the boundary is far from the horizon and positive when it is sufficiently close. In contrast, in the JT gravity black hole and cosmological systems with a reflecting boundary, the heat capacity is positive and negative, respectively, for any boundary position. We emphasize that we do not carry out any analysis of the relation between coarse-grained quantities such as heat capacity and fine-grained quantities such as entanglement entropy, so the observed parallel may simply be coincidental. Nevertheless, it seems worthwhile to further investigate this possible connection.

\begin{acknowledgments}
This work is  supported by the Russian Science Foundation (project 24-11-00039, Steklov Mathematical Institute). TR expresses gratitude to I.V. Volovich and I. Ya. Aref'eva for very valuable discussions regarding the setup and its mathematical subtleties. TR also thanks his fiancee Kristina L. for her support and care during the preparation of this work, which made it possible to bring the project to completion.
\end{acknowledgments}

\end{document}